 \documentclass[twocolumn,aps,prd,unsortedaddress,superscriptaddress,a4paper,nofootinbib,balancelastpage,preprintnumbers,showpacs,floatfix]{revtex4-1}

\usepackage{geometry}                
\usepackage{graphicx}
\usepackage{amssymb}
\usepackage{amsmath}
\usepackage{amsfonts}

\usepackage{dcolumn}

\def\ni{\noindent}
\def\be{\begin{equation}}
\def\ee{\end{equation}}

\def\Im{{\rm Im\,}}

\def\zu{\hat{\mathbf z}}
\def\kv{\mathbf k}
\def\pv{\mathbf p}

\def\T{_{\rm T}}

\def\bt{ b_{\rm T} }
\def\bl{ b_{\rm L} }
\def\kt{     { \kv_{\rm T}  }    }
\def\kpt{  \kv'_{\rm T} }
\def\ktkt{\kv^2_{\rm T}}
\def\kptkpt{{\kv'}^2_{\rm T}}

\def\pt{     { \pv_{\rm T}  }    }
\def\ptpt{\pv^2_{\rm T}}
\def\q{{\rm q}}

\def\h{{\rm h}}

\begin{document}

\title{Transverse momentum correlations of quarks in recursive jet models}


\author{X. Artru} 
\affiliation{Univ - Lyon,  CNRS/IN2P3, Universit\'e Lyon 1, Institut de Physique Nucl\'eaire de Lyon, 69622 Villeurbanne, France.}
\email{x.artru@ipnl.in2p3.fr}

\author{Z. Belghobsi}
\affiliation{Laboratoire de Physique Th\'eorique, Facult\'e des Sciences Exactes et Informatique
Universit\'e Mohammed Seddik Ben Yahia - Jijel \\
B.P. 98 ouled Aissa, 18000 Jijel, Algeria}
\email{belghobsi@univ-jijel.dz}

\author{E. Redouane-Salah} 
\affiliation{Universit\'e Mohamed Boudiaf de M'sila, D\'epartement de Physique, 28000 M'sila, Algeria.}

\begin{abstract}
In the symmetric string fragmentation recipe adopted by PYTHIA for jet simulations, the transverse momenta of successive quarks are uncorrelated. This is a simplification but has no theoretical reason. Transverse momentum correlations are naturally expected, for instance, in a covariant multiperipheral model of quark hadronization. We propose a simple recipe of string fragmentation which lead to such correlations. The definition of the jet axis and its relation with the primordial transverse momentum of the quark is also discussed. 
\end{abstract}

\maketitle 

\section{Introduction}

Most popular simulation models of quark jets are based on the recursive fragmentation:
\be  \label{cascade} 
\q_1 \to \h_1+\q_2\,, \cdots ,  \q_r \to  \h_r+\q_{r+1} \,, \cdots
\ee
$r$ is the \textit{rank} of hadron $\h_r$. To allow baryon production, a quark $\q_r$ can be replaced by an anti-diquark $(\bar\q\bar\q)_r$. At a step $\q\to \h+\q'$ we have $k=p+k'$, $k$ denoting the 4-momenta of  quark $\q$ and $p$ that of hadron $\h$. The momentum is shared following a \textit{splitting distribution}%
%
\be \label{mom-cons} 
dZ \, d^2\pt \, F_{h,q} (p,k)
\ee
with the normalization condition
\be  \label{normalize} 
\sum_h \int_0^1 dZ \int d^2\pt \, F_{h,q} (p,k) = 1.  
\ee
$\pt\equiv(p_x,p_y)$, $Z\equiv p^+/k^+$, $X^\pm \equiv X^0\pm X^z$.  
We place ourself in the final hadronic center-of-mass frame, define temporarily the jet axis to be along $\kv_1$ and take it as the $z$-axis.
$F$ depends on the flavor $q=u,d,s$ of the quark and the species $h$ of the hadron, whence the subscripts $h,q$.

Recursive fragmentation was first introduced in soft high-energy hadron-hadron collisions \cite{KRZ-PETERSON,FINKELSTEIN}, inspired by the multiperipheral model (MPM) \cite{AMATI} and its multi-Regge version \cite{CHEW}. These models yield the basic properties of jets in soft collisions, among which

$\bullet$ leading particle effect, 

$\bullet$ limiting fragmentation, 

$\bullet$
cutoff in $p\T$, 

$\bullet$
central plateau in rapidity, 

$\bullet$ short range order (in rapidity) 

$\bullet$
local compensation of charges \cite{KRZ-ch} and of 

\ \ transverse momenta \cite{KRZ-pt}.

\ni The quark diagram representing the hadronization of a high-energy quark-antiquark pair looks like a multiperipheral diagram where the exchanged particles are quarks instead of hadrons. It suggests \cite{is-pion-univ} a \emph{quark multiperipheral model} (QMPM) of hadronization (see figure 3). Such a model could explain why the properties listed above are also met in hard collisions. The \emph{leading particle effect} becomes \emph{quark charge retention} and the \emph{limiting fragmentation} becomes \emph{Feynman scaling}. 

The above properties can also be explained by the string fragmentation model (SFM) \cite{XA-GM,BOWLER,LUND}, in which a massive string stretching between the initial quark and antiquark (or diquark) decays into small mass strings representing hadrons or resonances. 
In fact the SFM can be considered as a special type of QMPM. In particular it can be treated in a recursive way  \cite{BOWLER,LUND}. The Lund group \cite{LUND} found that, in order to satisfy the following assumptions
\begin{enumerate}
\item symmetry under \emph{quark chain reversal}%
\footnote{called \emph{left-right symmetry} in \cite{LUND}. It comes in fact from charge conjugation symmetry.
},
\item invariance of $F_{h,q} (k,p)$ under a boost along $\zu$, a rotation about $\zu$ or a reflection about the $(z,x)$ or $(z,y)$ plane,
\item independence of $F_{h,q} (k,p)$ on $k^-$, for fixed $k^+$ and $\kt$,
\end{enumerate}
the splitting function should be of the form 
\begin{eqnarray}
\label{LSSF} 
 & & F_{h,q} (k,p) =  Z^{-1} \,\exp( -\bl {\epsilon^2}/{Z}) 
 \nonumber \\ 
&\times& \left(Z^{-1}-1\right)^{a_{q'}({\kpt}^2)} 
\left({Z}/{\epsilon^2} \right)^{a_{q}({\kv}\T^2)} 
\nonumber \\ 
 &\times& \ w_{q',h,q} (\kptkpt,{\pt}^2,\kt^2) / u_q(\kt^2)
\end{eqnarray}
This form is referred to as the \emph{Lund symmetric splitting function} (LSSF).  $\epsilon = (p^+ p^-)^{1/2} = (m_h^2+p\T^2)^{1/2}$ is the hadron \textit{transverse energy}. $w$ is symmetrical under $\{q,\ktkt\} \rightleftharpoons \{q',\kptkpt\}$ together with $h\to\bar h$.
$u_q(\kt^2)$ is fixed by Eq.(\ref{normalize}). Thus the input of the model consists in the functions $a_{q}({\kv}\T^2)$ and $w_{q',h,q}(\kptkpt,{\pt}^2,\kt^2)$. 

The behavior of the LSSF at $Z\to1$ resembles that of a QMPM model with ``reggeized'' quarks \cite{XA-CORRESP}.
Assumption 3 is related to a factorization property the SFM (see Eqs. 6-9 of \cite{BO-SO-2000} or Eqs. 3.10-3.11 of \cite{XA-CORRESP}): in figure 2, the hadronic states generated between two breaking points Q and Q' only depend on the vector $\overrightarrow{\rm QQ'}$. It comes from causality in the classical 1+1 dimensional string model. 

The $Z$-dependence of the LSSF is much constrained, particularly due to the assumption 3. However, there is a large freedom in the choice of the {functions} $a_{q}$ and $w_{q',h,q}$. 
In this paper we take $a_{q}({\kv}\T^2)$ = \textit{constant parameter}, like in the PYTHIA code, and consider several choices for $w$. We compare them concerning the transverse momentum correlations between successive hadrons. 
It is important to know these correlations, which do not depend on quark spin, to disentangle them from those coming from quark spin, which we will study in a future paper. 

The paper is organized as follows. Section II analyses the PYTHIA Monte-Carlo recipe for generating the transverse momenta, writes down the corresponding splitting function and discusses the predicted $(\pt,\pv'_{\rm T})$ correlations. Section III makes the comparison with a more natural choice of splitting function. Section IV starts from the most general QMPM and looks at a \emph{locally covariant} QMPM, built with a covariant vertex function.
Section V discusses the choice of the jet axis, more precisely the \emph{jet hyperplane}, and the effect of the primordial $\kt$. 
For simplicity, we will forget from now on the dependence on the quark flavors $q,q'$ and on the hadron species $h$. Thus the subscripts $q$, $q'$, and $h$ will be removed.  

\section{The PYTHIA splitting function}
In the widely used Monte-Carlo simulation code PYTHIA, $w(\kptkpt,{\pt}^2,\kt^2)$ is not given explicitly. Instead, the input is the function $u(\ktkt)$ of Eq.(\ref{LSSF}),  normalized to 
\be
\int d^2\kpt \, u(\kptkpt) = 1.
\ee
For instance, neglecting a large-$\kt$ tail,
\be \label{Gauss-u} 
u(\kptkpt) = (\bt/\pi) \,\exp(-\bt\kptkpt) \,.
\ee
 The splitting $\q\to \h+\q'$ is generated in two steps~:

\ni- 1) The subroutine PYPT draws $\kpt$ following the distribution $u(\kptkpt) \, d^2\kpt$.

\ni- 2) The subroutine PYZDIS draws  $Z$ following the distribution
\be \label{PYZ} 
N^{-1}(\epsilon^2) \, Z^{-1}dZ \, (1-Z)^a \, \exp( -\bl \,\epsilon^2/Z),
\ee
where $N^{-1}(\epsilon^2)$ is the normalization factor given by
\be \label{N} 
N(\epsilon^2) = \int Z^{-1}dZ \, (1-Z)^a \, \exp( -\bl \,\epsilon^2/Z).
\ee
Thus, the PYTHIA splitting function is
\begin{eqnarray}
\label{PYSF} 
F_{h,q} (k,p) =  N^{-1}(\epsilon^2) 
\, u(\kptkpt)
\nonumber\\
\times \ Z^{-1}\, (1-Z)^a\, \exp( -\bl \,\epsilon^2/Z) \,,
\end{eqnarray}
corresponding to 
\be \label{PYG} 
w(\kptkpt,{\pt}^2,\kt^2) = u(\ktkt) \, u(\kptkpt) \, \epsilon^{2a} / N(\epsilon^2).
\ee
Since $\kpt$ is drawn first and without reference to $\kt$, the $Z$-integrated splitting distribution is a function of $\kpt$ only. Thus \textit{there is no $(\kt,\kpt)$ correlation in PYTHIA after integration over $Z$}. In particular,  
$\langle \kt\cdot \kpt \rangle = 0$, 
from where  
\begin{eqnarray} \label{eq-pyt} 
\langle \pv^2_{1,{\rm T}} \rangle  = \langle \ktkt \rangle  
\,; \ 
\langle \pv^2_{r,{\rm T}} \ \rangle = 
2 \langle \ktkt \rangle \  \text{for } r>1 , 
\nonumber\\
\label{eq-pyt'} 
\langle \pt\cdot \pv'\T \rangle = - \langle \ktkt \rangle
\,; \quad
\langle \pt\cdot \pv''\T \rangle = 0 \quad\quad
\end{eqnarray}
for ranks $r'=r+1$ and $r''\ge r+2$. 
The compensation of transverse momenta is very local: it is achieved by the adjacent particles. Note that $\langle \ptpt \rangle$ of the first-rank hadron is half that of the other ones.  

\section{A splitting function with $(\kt,\kpt)$ correlation}

The presence of the factor $N^{-1}(\epsilon^2)$ in the function $w(\kptkpt,{\pt}^2,\kt^2)$ looks rather artificial and there is no physical reason to exclude a $(\kt,\kpt)$ correlation. Let us consider another input function
\begin{eqnarray} \label{LYG} 
& w_{q',h,q}(\kptkpt,{\pt}^2,\ktkt) = \epsilon^{2a}
\nonumber\\ 
& \times \exp\{-\bt(\ktkt+\kptkpt) + c\, \bl \epsilon^2 \}
\end{eqnarray}
where $c$ is a new parameter, the meaning of which is given later. We will take $c\in[0,1]$. The splitting function is then
\begin{eqnarray}
\label{LYSF} 
F_{h,q} (k,p) = \frac{1}{Z\, M(\ktkt)}\ \exp(-\bt \kptkpt) 
\nonumber\\
(1-Z)^a\,  
\exp\left\{-\bl\epsilon^2\left(Z^{-1}-c\right)\right\},  
\end{eqnarray}
$1/M$ being the normalization factor given by
\begin{eqnarray}
M(\ktkt) &=& \int d^2\kpt \, \exp(-\bt \kptkpt)  
\nonumber\\
 &\times& N(\epsilon^2) \, \exp(c\,\bl\epsilon^2) \,,
\end{eqnarray}
with $M(\ktkt)  \, \exp(-\bt \ktkt)=u(\ktkt)$ of Eq.(\ref{LSSF}).

Figure 1 compares the shapes of the splitting function (\ref{LYSF}) and the PYTHIA one, Eqs.(\ref{Gauss-u},\ref{N},\ref{PYSF}). The main difference is that the barycenter of (\ref{LYSF}) is on the side of $\kt$, whereas it stays at $\kpt=0$ for Eq.(\ref{PYSF}). 
The $(\kt,\kpt)$ correlation in Eq.(\ref{LYSF}) is due to the last exponential which is a decreasing function of $|\pt|=|\kpt-\kt|$ for all $Z$. In Eq.(\ref{PYSF}) this effect is exactly compensated by the factor $1/N(\epsilon^2)$ when we average over $Z$.

\begin{figure} 
\includegraphics*[angle=0, width=72mm, bb=35 0 440 285]{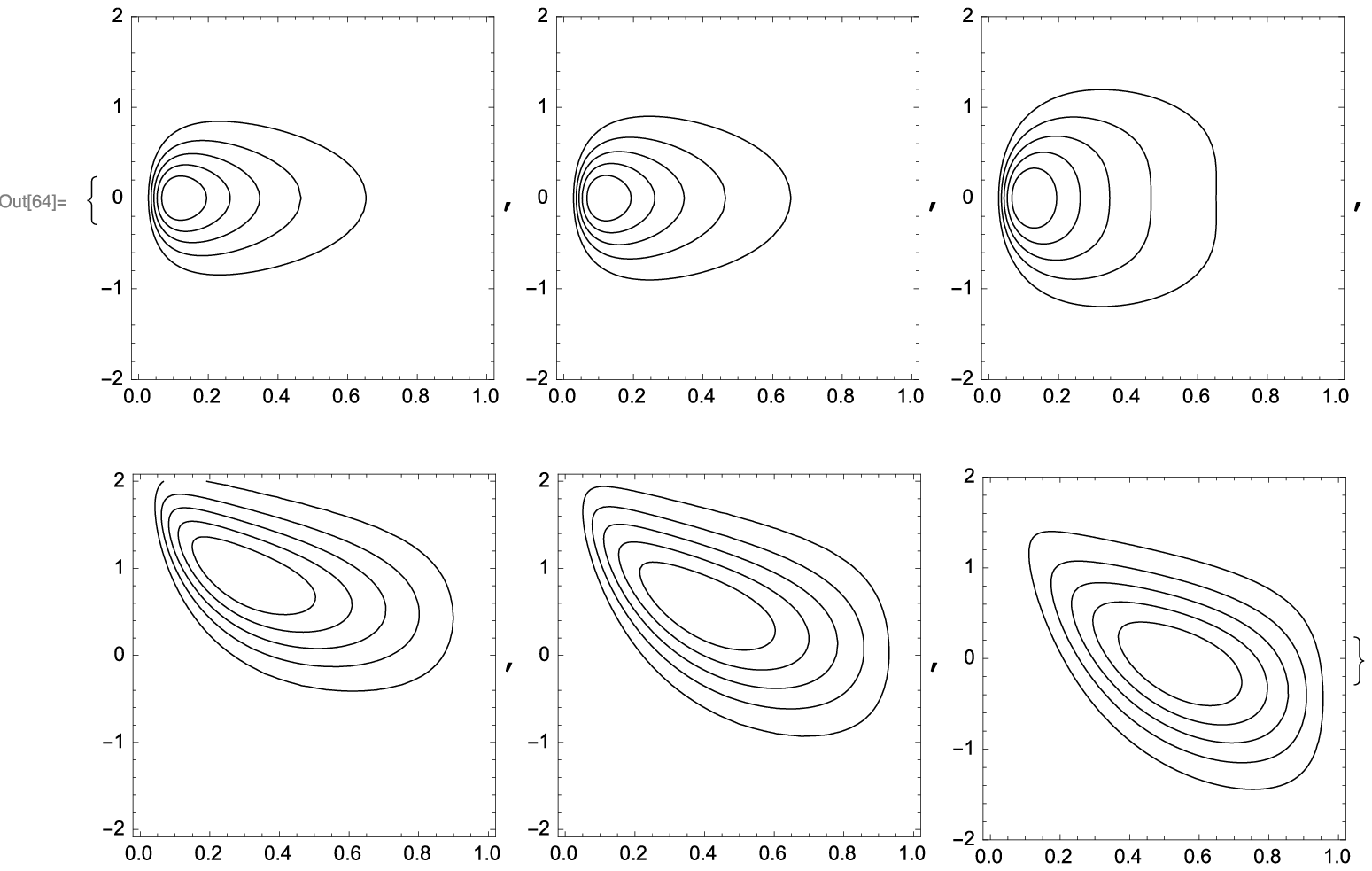}  
  \caption{\footnotesize
Comparison of the splitting functions of Eq.(\ref{LYSF}) and Eqs.(\ref{Gauss-u},\ref{N},\ref{PYSF}) in the $(Z,k'_x)$ plane, for $k'_y=k_y=0$. Vertical scales: $k'_x$ in fm$^{-1}$.  
Upper figures: $k_x=0$ ;   lower figures: $k_x$ = 2 fm$^{-1}$.
Left figures: Eq.(\ref{LYSF}) with $c=0$; middle figures: Eq.(\ref{LYSF}) with $c=1$; right figures: Eqs.(\ref{Gauss-u},\ref{N},\ref{PYSF}). The other parameters are $a=1$, $\bt = 1$ fm$^2$, $\bl = 0.25 $ fm$^2$, $m_h=m_\pi=0.7$ fm$^{-1}$.
}
\end{figure}

The  Monte-Carlo drawing of $Z$ and $\kpt$ from Eq.(\ref{LYSF}) proceeds in two steps~:

\ni 1)  draw $Z$ according to the $\kpt$-integrated distribution
\be \label{tirZ} 
 \frac{\pi\, dZ \,(1-Z)^a}{ Z [\bt+b_Z]} \,\exp\left\{
-{b_Z \, m_h^2}- \frac{\kt^2}{\bt^{-1}+b^{-1}_Z}\right\}
\ee
with $b_Z\equiv\bl \, (Z^{-1} -c)$,

\ni 2) draw $\kpt$ according to the fixed-$Z$ distribution 
\be  \label{tirpT} 
\frac{\bt+b_Z}{\pi} \,
\exp\left\{-[\bt+b_Z] \left(\kpt - \frac{\kt}{1+\bt/b_Z}\right)^2 \right\},
\ee
which is a translated Gaussian. At fixed $Z$,
\be \label{kpt-moyen} 
\langle\kpt\rangle = \frac{\kt}{1+\bt/b_Z} \equiv \lambda(Z)\,\kt 
\ee
with $\lambda(Z)\in[0,1]$, $\,\lambda(0) = 1$ and $d\lambda/dZ<0$. 
After integration over $Z$ we have, for $r\ge2$,
\be \label{ineq-lyo} 
\langle \kt \cdot \kpt \rangle >0  \,, \quad
\langle \ptpt \rangle < 2 \,\langle \ktkt \rangle 
 \,.
\ee
The $\kt$ correlation spreads over several ranks, while decreasing: 
\be  \label{ineq-ly1} 
\langle \kt \cdot \kpt \rangle >
\langle\kv_{r,{\rm T}}\cdot \kv_{r+2,{\rm T}}\rangle > \cdots >  0\,,
\ee
from where 
\begin{eqnarray} \label{ineq-ly2} 
 &- \langle \ktkt \rangle + \langle \kt \cdot \kpt \rangle 
<\langle\pv_{r,{\rm T}}\cdot \pv_{r+1,{\rm T}}\rangle 
\nonumber\\
& < \langle\pv_{r,{\rm T}}\cdot \pv_{r+2,{\rm T}}\rangle < \cdots < 0 \,.
\end{eqnarray}
Equations (\ref{ineq-lyo}-\ref{ineq-ly2}) are at variance with Eq.(\ref{eq-pyt}). The compensation of transverse momenta needs more than the adjacent hadrons. 
Table 1 shows results from Monte-Carlo simulations with our splitting function, Eq.(\ref{LYSF}) and the PYTHIA splitting function, Eqs.(\ref{Gauss-u},\ref{N},\ref{PYSF}). 
In the first column $\langle \ln(1\!-\!Z)\rangle$ is the average rapidity step per splitting. It gives the particle density in rapidity space, 
\be
dN/dY = - \langle \ln(1\!-\!Z)\rangle^{-1} \,.
\ee
Note that an increase of $c$ brings our splitting function closer to the PYTHIA one.

Both PYTHIA (see Eq. \ref{eq-pyt'}) and the present model predict %
\begin{eqnarray} \label{pTrang1} 
\langle \pv^2_{1,{\rm T}} \rangle  < \langle \pv^2_{r,{\rm T}} \ \rangle\  \text{for } r>1 \,,
\end{eqnarray}
%
because $\kv_{1,{\rm T}}=0$. This is confirmed by computer simulations by A. Kerbizi \cite{ALBI} and us. However this result does not take into account the \emph{primordial} transverse momentum (see section V).

An equation similar to (\ref{kpt-moyen}) has been proposed in \cite{AND-GUST-SAM}, but with a constant $\lambda$.
In their notations, $k_i$, $p_{i-1}$ and $\gamma$ correspond to our $p$, $k$ and $1-\lambda$ respectively. 

The inclusion of the quark spin degree of freedom may reverse the inequalities (\ref{ineq-lyo}). Indeed, the $^3P_0$ model would predict $\langle \kt \cdot \kpt \rangle <0$ and  $\langle \ptpt \rangle > 2 \,\langle \ktkt \rangle$ for $\h$ = pseudoscalar meson \cite{DUBNA13} if the natural $(\kt,\kpt)$ correlations studied here were absent. This is a ``hidden spin'' effect, independent of the initial quark polarization. 
\begin{table}
\caption{\footnotesize{
Results from Monte-Carlo simulations. First two lines: our splitting function (\ref{LYSF}) with $c$ = 0 and 1. Third line: PYTHIA splitting function, Eqs.(\ref{Gauss-u},\ref{N},\ref{PYSF}). The statistics comprises 100 jet, each one containing 1000 particles (no lower cutoff in energy was imposed), so the observables are those of the rapidity plateau. $\pt$ and $\kt$ are in GeV units. Primed quantities refer to a particle of adjacent rank. The parameters are the same as in figure 1.
}}  
\begin{center}
\begin{tabular}{l|lllll|}
\hline
\\ 
model & $\langle \ln(1\!-\!Z)\rangle$ & $\langle\ktkt\rangle^{\frac{1}{2}}$ & $\frac{\langle\kt\cdot\kpt\rangle }{ \langle\ktkt\rangle}$ 
& $\langle\ptpt\rangle^{\frac{1}{2}}$ & $\frac{\langle\pt\cdot\pv'_{\rm T}\rangle}{ \langle\ptpt\rangle}$
\\ [1ex]
\hline \\
 $c=0$ &\ -0.562 & 0.167 & 0.430 &  0.179 & -0.287
\\[1ex]
 $c=1$ &\ -0.598 & 0.176 & 0.302 & 0.207 & -0.352
\\[1ex]
PYT.  &\ -0.998 & 0.200 & $ 0 $ & 0.283 & - 1/2
\\[1.5ex]
\hline
\end{tabular}
\end{center}
\end{table}

\medskip
\ni\textbf{Meaning of the parameter {\boldmath$c$}.}
The product of the factors $\exp\left( -\bl {\epsilon^2}/{Z} \right)$ of Eq.(\ref{LSSF}) for the successive splittings is equal to $\exp(- {\cal P}{\cal A})$, where ${\cal A}$ is the total area shown in Fig. 2, ${\cal P}=-2\,\Im(\kappa_C)= 2\,|\kappa_C|^2\bl$ and $\kappa_C$ is the complex string tension. ${\cal A}$ is the space-time area of the string world sheet which is still unaffected by the splittings and ${\cal P}$ can be called ``string fragility''. 
One may ask why the light-grey rectangles like ${\rm C_2Q_2H_2Q_3}$ should be included in ${\cal A}$. Indeed, if $\h_2$ is a stable hadron, a string cutting in this rectangle is kinematically forbidden. Therefore it seems theoretically preferable not to include such rectangles in ${\cal A}$, keeping only the dark-gray area. It amounts to replace $\exp\left( -\bl {\epsilon^2}/{Z} \right)$ by $\exp\left( -\bl {\epsilon^2}(1/Z-1) \right)$. To keep the two possibilities open, we introduced the parameter $c$ in Eqs.(\ref{LYG}-\ref{LYSF}). With $c=0$ the rectangles are included, with $c=1$ they are excluded. We also allow admit $c\in[0,1]$. Note that the extra factor $\exp\left(c\,\bl {\epsilon^2} \right)$ does not break the symmetry under quark chain reversal. It can indeed be absorbed in the function $w_{q',h,q} (\kptkpt,{\pt}^2,\kt^2)$ of Eq.(\ref{LSSF}), as we did in Eq.(\ref{LYG}).
%
\begin{figure} %
  \centering 
\includegraphics*[angle=90, width=70mm, bb=120 160 485 705]{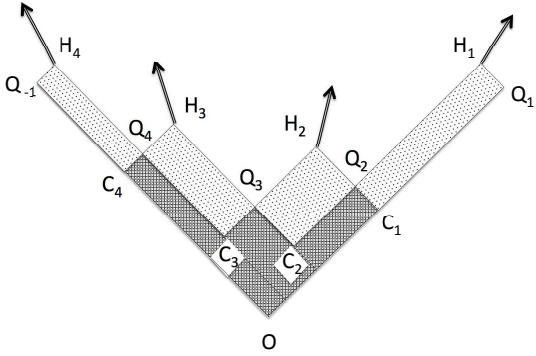}  
  \caption{\footnotesize
Space-time picture of the string fragmentation model. The string breaks at points ${\rm Q_2\cdots Q_4}$. Hadrons liberated at points ${\rm H_1\cdots H_4}$ (choice $c=0$), but they can be considered as really ``born'' at points ${\rm C_1\cdots C_4}$ (choice $c=1$).}
\end{figure}

\section{Local versus global covariance}

\subsection{The general quark multiperipheral model.}
Let us consider the hadronization of a quark-antiquark pair $\bar\q_{-1} \q_1$ of high center-of-mass energy squared, $s={k_{-1}^-k_1^+}$. The most general QMPM for this reaction can be defined by the distribution of $\q\to\h+\q'$ vertices in the 8-dimensional $\{k,k'\}$ space, 
\begin{eqnarray} \label{vertex-dis} 
& &d^4k\, d^4k' \, 2 \delta(p^2-m_h^2) \ {\cal W}_{q',h,q}(k,k') 
\nonumber \\ 
& &\times \ { \sigma_{\bar\q_{-1},\q'} \!\left(\bar k_{-1},k'\right) \, 
\sigma_{\bar\q,\q_1} \! \left(\bar k,k_1\right) }
\nonumber \\ 
& &\times \ 1/{  \sigma_{\bar\q_{-1},\q_1} \! \left(\bar k_{-1},k_1\right) }  \,.
\end{eqnarray}
This equation is suggested by figure 3. $\bar k\equiv -k$ ; $\,{\cal W}_{q',h,q}(k,k')$ is a vertex function which includes the adjacent quark propagators. It is symmetrical under $\{q,k\} \rightleftharpoons \{q',k'\}$, $h\to\bar h$ ;
\be
 \sigma_{\overline {\rm B},{\rm A}}\left(\bar k_{\rm B},k_{\rm A}\right) \equiv \sum_{X} \sigma\left\{\overline {\rm B}+{\rm A} \to X\right\} \, 
\ee
is the total ``cross section'' between a quark A  of the chain and the line-reverse of a quark B of higher rank. At large $\left(\bar k_{\rm B}+k_{\rm A}\right)^2$ the QMPM predicts the Regge behavior
\be
  \sigma_{\overline {\rm B},{\rm A}}\left(\bar k_{\rm B},k_{\rm A}\right) 
 \simeq \beta_{\rm B}(k_{\rm B}) \, 
 \beta_{\rm A}(k_{\rm A}) \, \left| k_{\rm B}^-k_{\rm A}^+\right|^\alpha .
\ee
We will omit the Regge residues $\beta$ because they can be incorporated in ${\cal W}_{q',h,q}(k,k')$. The splitting function is 
\be
F_{h,q} (p,k) = {\cal N}^{-1}_q(k) \,  Z^{-1} (1-Z)^\alpha \, {\cal W}_{q',h,q}(k,k')  
\ee
with
\be  \label{U} 
{\cal N}_q(k) = \sum_h \int_0^1 \! dZ \frac{(1-Z)^\alpha}{Z}  \int \! d^2\pt \, {\cal W}_{q',h,q}(k,k') 
\ee
and the recurrent quark density in the central region (large $|k^-k_1^+|$ and $|k_{-1}^-k^+|$) is 
\be
{\cal U}_q(k) = |k^+k^-|^\alpha \, {\cal N}_q(k) \,.
\ee
The LSSF (Eq. \ref{LSSF}) is obtained with $\alpha=0$ and the vertex function 
\begin{eqnarray} \label{SLW} 
{\cal W}_{q',h,q}(k',k) =  \left(\frac{{k'}^+}{p^+}\right)^{a_{q'}(\kptkpt)} \left|\frac{k^-}{p^-}\right|^{a_{q}(\ktkt)}  
\nonumber \\
 \times \exp (\bl \,{k'}^-k^+) \, w_{q',h,q} (\kptkpt,{\pt}^2,\kt^2) \quad\quad
\end{eqnarray}
(equivalent to Eq.(3.2) of \cite{XA-CORRESP}). 
${\cal W}$ of Eq.(\ref{SLW}) is not invariant under the full Lorentz group, due to assumption 3 of Section I, so we say that the model is not \emph{locally} covariant. In fact, the quark pairs $\q_r\bar\q_r$ are not created in vacuum but in the string thickness, which is an oriented medium. 
The SFM is however \emph{globally} covariant if the initial quarks $\bar\q_{-1}$ and $\q_1$ are generated in a covariant way. 

\subsection{A locally covariant model.}
Let us consider a QMPM based on the old-fashioned mutiperipheral model \cite{AMATI} for spinless particles, with covariant vertices and propagators. Assumption 3 is abandoned and
\be
{\cal W}_{q',h,q}(k,k') = W_{q',h,q}(k^2,{k'}^2) \,, 
\ee
$W_{q',h,q}(k^2,{k'}^2)$ decreasing rapidly at large $|{k}^2|$ and $|{k'}^2|$. This model predicts a $(\kt,\kpt)$ correlation. Indeed, the kinematical relation
\be \label{kp2} 
{k'}^2 = (1-Z)(k^2-m_h^2/Z) - \pv_{\perp \kv}^2/Z \,,
\ee
with $\pv_{\perp \kv} \equiv \pt-Z\,\kt$, leads to 
\be \label{kpt-moyen-bis} 
\langle \pt \rangle  = Z\,\kt  \,, \quad \langle \kpt \rangle  = (1-Z)\,\kt  \,.
\ee
This relation is to be compared to Eq.(\ref{kpt-moyen}). It comes from the possibility to re-orient the $z$-axis after each splitting along the new quark momentum. It also assumed in the model of Ref. \cite{MATEVOSYAN}.

Equation (\ref{kpt-moyen-bis}) also occurs for $c\!=\!1$ and $\bl\!=\!\bt$ in the model of Section III, according to Eq.(\ref{kpt-moyen}). For these parameters, Eqs. (\ref{LYG}) and (\ref{SLW}) give
\begin{eqnarray} \label{Wcov}
 & {\cal W}_{q',h,q}(k,k') = |{k'}^+k^-|^a \quad
\nonumber \\
 & \times \exp\{\bt ({k'}^2 + k^2 -{k'}^+k^-)\} \,.
\end{eqnarray}
Owing to ${k'}^+k^- = (1-Z) \, k^+k^-$, the $k'$ dependance in Eq.(\ref{Wcov}) at fixed $Z$ and $k$  is only through the covariant variable ${k'}^2$. However the model is not fully covariant since a redefinition of the $z$ axis changes $ k^+k^-$, therefore changes the $Z$ distribution.

\section{Jet hyperplane and primordial $\kt$}

The $\pt$'s, $\kt$'s and their correlations depend on the precise definition of the jet axis, represented by the unit vector $\zu$.  Indeed, a change $\Delta \zu$ of $\zu$ induces a change $\Delta\pt\simeq - |\pv|\, \Delta \zu$ of $\pt$.
In section I we temporarily defined the jet axis to be along $\kv_1$ in the final hadronic center-of-mass frame. 
This definition involves  two 4-vectors, $k_1$ and the total hadronic 4-momentum $P$. 
The jet properties listed in Section I are invariant under a boost along the axis, \emph{i.e.}, under a Lorentz transformation in the 2-D hyperplane $(k_1,P)$ spanned by $k_1$ and $P$, so one should rather speak of a \emph{jet hyperplane}. This hyperplane can equally be defined by $k_1$ and $\bar k_{-1} = - k_{-1}=P-k_1$. In the SFM, it contains the string world sheet. 
In deep inelastic lepton scattering (DIS), $\bar k_{-1}$ is the 4-momentum of the target remnant. 

The choice of the $(k_1,P)$ hyperplane is somewhat arbitrary: in DIS, one may prefer the hyperplane $(p_{\rm target},k_1)$. From the experimental point of view, only the $(p_{\rm target},p_{\gamma*})$ hyperplane is well defined ($\gamma*$ is the virtual photon). Besides, in any jet producing reaction (DIS, high-$\pt$ jets, $e^+e^-$ annihilation, etc;) $k_1$ is not well defined theoretically. It is an internal momentum in a loop diagram (see figure 1 of \cite{DUBNA13}) and must be integrated over, therefore the cross section is a double integral: in $k_1$ for the amplitude  and in $k'_1$ for the complex conjugate amplitude. The loop topology is imposed by confinement. All recursive simulation models are classical in that they ignore the difference $k'_1-k_1$. 

\medskip
\ni\textbf{\boldmath Handling the primordial $\kt$.}
In DIS there is a \emph{primordial} transverse momentum $\kt_{\rm prim}$ with respect to the $(p_{\rm target},p_{\gamma*})$ hyperplane. In simulations, it is generated randomly and can be taken into account in two ways:

- a) choose $(p_{\rm target},p_{\gamma*})$ as jet hyperplane, from where $\kt_1=\kt_{\rm prim}$.

- b) choose a  jet hyperplane containing $k_1$, from where $\kt_1=0$.

\ni These two choices are equivalent for a locally covariant model, but not for the string fragmentation model, which makes the choice b). Choice a) would mean that the quark $\q_1$ is not drawing a rectilinear string. The other side of the string, however, is attached to the target remnant which has a transverse size. For this side, choice a) is not worse than b).

The inequality (\ref{pTrang1}) holds if $\pt$ is defined relative to $\kv_1$. Relative to $\pv_{\gamma*}$, it
may be attenuated or even reversed, due to the primordial  $\kt$, which adds  $(p^+/k_1^+)^2\langle {\kv\T^2}\rangle_{\rm prim} $ to $\langle \ptpt \rangle$. 

%
\begin{figure} %
  \centering   
\includegraphics*[angle=90, width=70mm, bb=220 250 370 610]{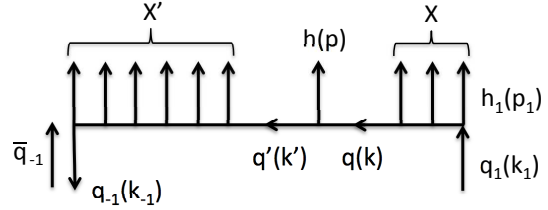}  
  \caption{\footnotesize
multiperipheral diagram illustrating Eq.(\ref{vertex-dis}).}
\end{figure}

\section{Conclusion}

We have shown that a $(\kt,\kpt)$ correlation with $\kt\!\cdot\!\kpt>0$  is naturally expected in a quark multiperipheral model of hadronization, whereas PYTHIA assumes no correlation.  We have underlined the difference between the string fragmentation models based on assumption 3 of Section I and a locally covariant QMPM, where this assumption  is replaced by the full covariance of the vertex function. In the latter,  the jet axis can be redefined after each splitting so that the new $\kt$ is vanishing. It implies the $(\kt,\kpt)$ correlation (\ref{kpt-moyen-bis}). 
The re-orientation of the jet axis cannot be done in the string fragmentation model  because the string is supposed to maintain its direction. 

We have also pointed out theoretical ambiguities in defining the jet axis and compared two ways of handling the primordial transverse momentum in DIS. 
It would be interesting to find an experimental trace of the inequality (\ref{pTrang1}) (where $\pt$ is defined relative to $\kv_1$), in spite of the $\kt_{\rm prim}$ effect. 

We must keep in mind that the $\pt$ correlations studied here should be a background under the resonance effects, the Bose-Einstein correlations and the spin-induced correlations like the di-hadron Collins asymmetry
observed by the COMPASS collaboration \cite{COMPASS}
and the ``hidden spin'' effects suggested in \cite{DUBNA13}.  

\subsection*{Acknowledgment}
We thank Prof. F. Bradamante, Prof. A. Martin and A. Kerbizi for their continuous interest in this work.

\end{document}